# Low-resistivity nitrogen-doped p-type $Cu_2O$ thin films enabled by millisecond flash lamp annealing

Jan Koloros[a], Pavel Baroch[a], Thomas Preußner[b], Matthias Fahland[b], Michaela Červená[a], Jiří Rezek[a]*

[a] Department of Physics and NTIS, European Centre of Excellence, University of West Bohemia in Pilsen, Univerzitní 8, 301 00 Pilsen, Czech Republic

[b] Fraunhofer Institute for Electron Beam and Plasma Technology – FEP, Winterbergstr. 28, 01277 Dresden, Germany

*Corresponding author, Tel.: +420 377632269, E-mail address: jrezek@fav.zcu.cz

Abstract

Flash lamp annealing (FLA) provides millisecond-scale thermal processing, but its effects on p-type $Cu_2O$ and nitrogen-related defects remain poorly understood. Reactively sputtered $Cu_2O$:N films with different nitrogen content were deposited using reactive high-power impulse magnetron sputtering and exposed to a single 1.9 ms FLA pulse at 4.9–11.7 J $cm^{-2}$. Their compositional, morphological, structural, vibrational, electrical, and optical responses were evaluated. WDS showed no statistically significant change in total elemental composition, and XRD confirmed retention of cubic $Cu_2O$. Nitrogen-containing films exhibited surface coarsening, shifts of the $Cu_2O$ reflections, and non-monotonic changes in the Raman band assigned to molecular $N_2$. Nitrogen incorporation substantially reduced the as-deposited resistivity. The very low value of $4.5 \times 10^{-2}$ Ω cm was obtained after FLA at 4.9 J $cm^{-2}$, whereas higher energy densities markedly increased resistivity. Hall measurements showed increasing mobility but decreasing hole concentration. At high energy densities, the optical band gap of nitrogen-rich films widened. The results define a narrow low-energy processing window with a positive effect on electrical properties, whereas high-energy FLA modifies the structure and optical absorption edge but degrades electrical conductivity.

## 1. Introduction

Driven by rapid technological evolution, recent research has intensively focused on enhancing the cost-effectiveness and efficiency of transparent conductive oxides (TCO). TCO represent a class of materials that combine low electrical resistivity and high optical transparency. These properties are necessary for a wide variety of applications, such as thin-film solar cells, thin-film transistors, smart windows, photodetectors, and transparent electronic devices [1–6]. Among various alternatives, $Cu_2O$-based materials have received attention for their favourable balance of abundance and performance. These films have several applications, such as sensors, transistors, photocatalysts, and photovoltaic cells [7–12].

However, p-type TCO generally suffer from poorer electrical properties compared to their n-type counterparts. The synthesis of p-type TCO presents a challenge due to the intrinsic localized nature of the valence band, primarily derived from oxygen 2p orbitals. This causes lower mobility of holes compared to electrons [1]. Having comparable performance is necessary to fabricate electronic devices based on transparent p-n junctions. [1,13].

To enhance the electrical conductivity of transparent conductive $Cu_2O$ thin films, many strategies are commonly used. Trolio et al. used H and N doping of $Cu_2O$ films to reduce the electrical resistivity to 0.16 Ωcm [14]. Hsiao et al. utilized electrochemical deposition to grow $Cu_2O$ films, reporting good epitaxial growth and superior crystallinity [15]. In our recent work, we have used post-deposition laser processing and nitrogen doping. In the first work, we have applied laser thermal annealing (LTA) to sputtered $Cu_2O$ films, reporting an increase in hole mobility to $24\ cm^2V^{-1}s^{-1}$ after the treatment [16]. In the latter one we reported that nitrogen doping via reactive high-power impulse magnetron sputtering (r-HiPIMS) enables a significant reduction in resistivity to approximately $5 \times 10^{-2}$ Ωcm. However, it reduces the hole mobility [17]. Alternatively, conductivity can be optimized through techniques such as deposition at elevated temperatures, post-deposition annealing, all of which have been shown to enhance hole mobility and reduce electrical resistivity in $Cu_2O$.

Recently, flash lamp annealing (FLA) has attracted interest as an advanced thermal treatment. FLA is efficiently utilized to improve the homogeneity and crystallinity of thin films. Beyond its scalability and rapid processing speed, a key advantage of FLA is that the millisecond-scale annealing duration eliminates the need for a controlled atmosphere. To date, the evaluation of FLA has been predominantly restricted to n-type transparent conductive layers, with numerous studies dedicated to ITO, ZnO, and AZO [18–20]. For AZO thin films, FLA leads to an order-of-magnitude increase in electron mobility, while the electron concentration rises by up to two orders of magnitude. Consequently, the overall sheet resistance decreases by several orders of magnitude [18]. According to Skorupa et al., utilizing FLA decreased the resistivity of the ZnO:Al films by a factor of two, down to $1.0 \times 10^{-3}$ Ωcm [19]. Kim et al. showed that utilizing FLA can reduce the electrical resistivity of ITO films by up to 30% while improving their mobility, achieving results comparable to annealing the samples for 1 hour at 200–300 °C [20].

Conversely, the influence of FLA on p-type thin films remains poorly understood, and relevant literature is scarce. One study exploring FLA on $Cu_2ZnSnS$ demonstrated enhanced crystallinity post-treatment [21]. Nevertheless, owing to the low initial conductivity of the films, the correlation between FLA and electrical transport properties was not investigated. To date, no studies have been published investigating the effects of FLA as a post-deposition treatment on $Cu_2O$-based thin films.

This manuscript investigates the effect of FLA on the properties of $Cu_2O$:N thin films. Our results demonstrate that FLA represents an effective approach to controlling key transport properties, specifically hole mobility and concentration. Additionally, this treatment enables a reduction in electrical resistivity down to $4.5 \times 10^{-2}$ Ωcm and seems to be promising, low-cost way for the tailoring of p-type TCOs.

## 2. Experimental setup

### 2.1. Film preparation

Reactive high-power impulse magnetron sputtering was employed to deposit thin Cu-O-N films using a 99.99% pure copper target (100 mm diameter, 6 mm thickness) at the substrate holder-target distance of 10 cm. The chamber was pumped down to a base pressure of approximately $6\times10^{-4}$ Pa. The deposition apparatus is identical to that used in our two previous studies [16,17] and is illustrated in **Fig. 1**. The gases' mass flow settings involved two stages: First, stabilizing the Ar + $N_2$ pressure at 0.5 Pa while adjusting the nitrogen fraction ($f_{N2}$). Second, the oxygen was introduced at a constant partial pressure of 0.27 Pa. The nitrogen fraction ($f_{N2}$) is given as the mass flow rate ratio:

$$f_{N2} = \frac{mass\ flow\ of\ N_2}{mass\ flow\ of (N_2 + Ar)}) \qquad (1)$$

A SPIK2000USB_S (Melec GmbH) power supply providing rectangular voltage pulses ($t_{on}$ = 100 μs, duty cycle ≈ 0.49 %) was used. An average target power of 500 W was kept constant for all deposition, resulting in the pulse-averaged target power density, $S_{da}$, of 1300 Wcm$^{-2}$. Soda-lime glass (18 × 18 mm$^2$) and Si (100) wafers (10 × 10 mm$^2$) were used as substrates. Before deposition and cleaning, laser-scribed grooves were patterned on the glass substrates to facilitate easier and more precise post-deposition cleaving into four pieces of 9 × 9 mm$^2$. The substrates were cleaned in isopropyl alcohol and distilled water (10 min for each process). The substrate holder was heated to 200 °C during the deposition. The deposition time was kept constant for all depositions (120 s), resulting in films with a thickness of ~250 nm (±5 %).

## 2.2. Flash lamp Annealing

Flash Lamp Annealing (FLA) has been used for post-deposition annealing of the films. By transferring photonic energy by a short and intense light pulse to the samples, the annealing of surface in the millisecond range is enabled. Due to the rapid heating rate, the surface region including the thin film can reach very high temperatures, while the bulk of the glass substrate remains "cold". The total level of the achievable temperature as well as its distribution into the substrate depend on thermal materials properties and parameters of the light pulse. In section 2.4. a numerical approach for quantitative temperature calculation is proposed.

A custom-made FLA assembly at Fraunhofer FEP (ROVAK GmbH) was used. The FLA-processed $Cu_2O$:N samples had a size of 9 x 9 $mm^2$ and were treated by a single flash at ambient atmosphere (**Fig. 1**). The FWHM of the light pulse has been set to 1.9 ms. The total energy density, $E_{density}$, delivered ranged from 4.9 to 11.7 $Jcm^{-2}$.

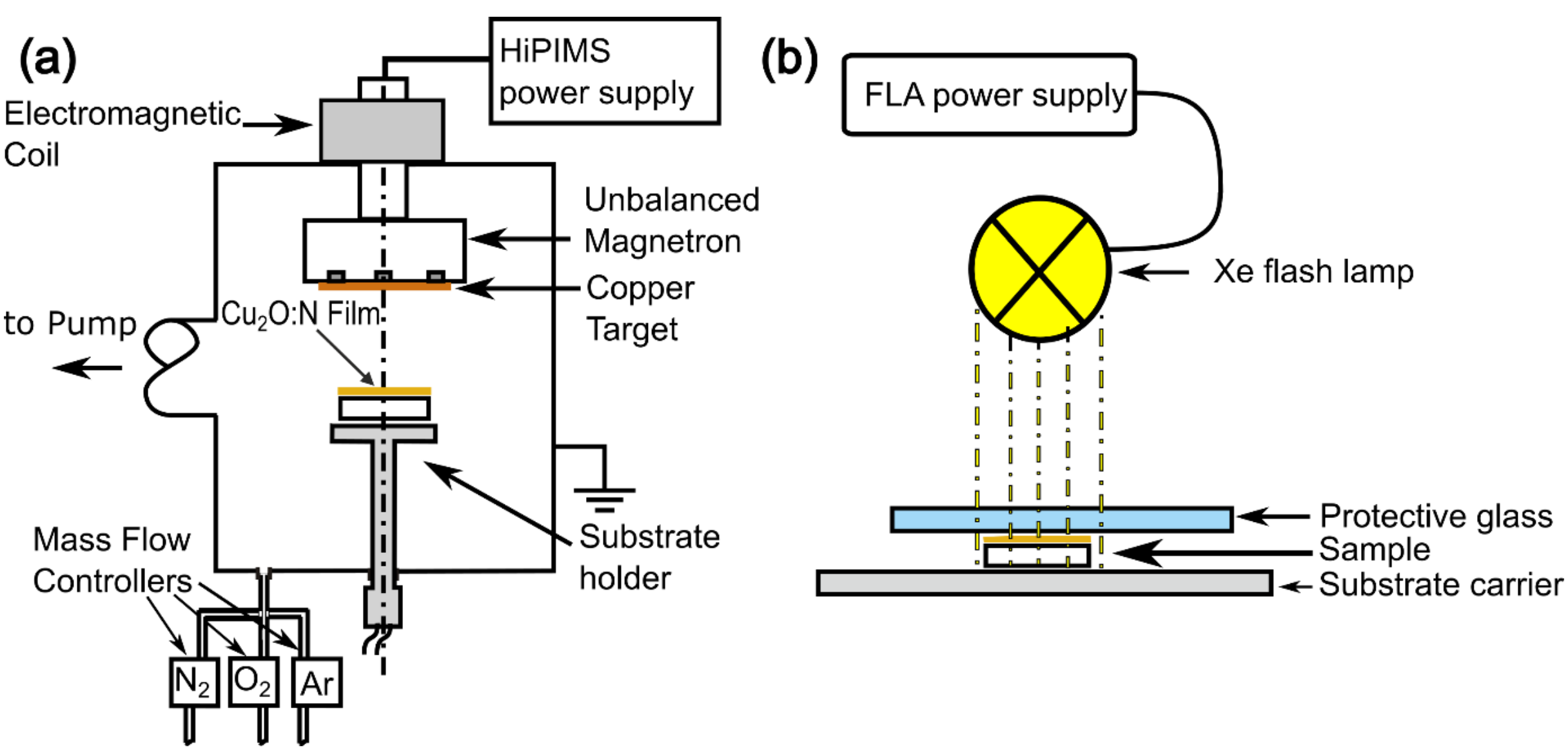

Fig. 1. (a) A schematic picture of a deposition system for deposition of $Cu_2O$:N thin films, (b) a schematic picture of a flash lamp thermal annealing apparatus.

## 2.3. Film characterization

Raman spectroscopy (Horiba Jobin Yvon LABRAM HR Evolution) equipped with a 532 nm green laser was used to detect and characterize molecular nitrogen features around a wavenumber of 2250 $cm^{-1}$. The crystalline structure of the films was characterized by Grazing Incidence X-ray Diffraction (GIXRD) using a multifunctional Rigaku SmartLab diffractometer equipped with a high-power 9 kW rotating anode X-ray generator and a primary Johansson $K\alpha 1$ monochromator. Wave-dispersive spectroscopy (WDS) inside a scanning electron microscope (SEM, Hitachi SU-70) was employed to determine the elemental composition of the Cu-O-N films. The SEM was also utilized for top-view surface imaging, operating at accelerating voltages ranging from 2 to 15 kV.

The electrical resistivity was measured at room temperature by the standard four-probe method using tungsten probes. The dimensions of the square samples were $9 \times 9$ $mm^2$, the inter-probe distance was 1 mm, and the probe tip radius was 150 μm. The carrier concentration and mobility were evaluated via a Hall Measurement System (MMR Technologies) using the Van der Pauw configuration. To improve the electrical contact between the sample and the gold probes of the Hall system, 50 nm thick gold contacts were sputter-deposited onto the corners of the $9 \times 9$ $mm^2$ samples.

The optical band gap, $E_g$, of the thin films was determined using the Tauc plot method. Transmittance $T$ and reflectance $R$ spectra were measured in the 250–2500 nm range using an Agilent Technologies Cary 7000 spectrophotometer, with the reflectance recorded at a 14° angle of incidence. The absorption coefficient α was calculated from the measured $T$, $R$, and film thickness [17]. For the direct optical band gap, the Tauc relation was applied: $(\alpha h\nu)^2 = A(h\nu - Eg)$, where $h\nu$ is the photon energy and $A$ is a constant. The value of $E_g$ was then determined by plotting $(\alpha h\nu)^2$ against $h\nu$ and extrapolating the linear region of the curve to the x-axis intercept.

## 2.4. Thermal models of FLA annealing

Thermal models for flash lamp annealing have been reported by various authors [22,23]. They are based on the numerical solution of the heat conduction equation. The one-dimensional heat conduction equation can be written as

$$\frac{\partial}{\partial t}\left(\rho c_p T\right) = \frac{\partial}{\partial x}\left(\lambda \frac{\partial}{\partial x} T\right) + Q_{Vol} \tag{2}$$

where $\rho$, $c_p$ , $\lambda$ and $Q_{Vol}$ are the density, specific heat capacity, thermal conductivity and the absorbed radiation power per volume, respectively. This is simplified under the assumption that there are no mass flows and that the medium is homogeneous.

$$\frac{\partial}{\partial t} T(x,t) = \alpha \frac{\partial^2}{\partial x^2} T(x,t) + \frac{Q_{Vol}}{\rho\, c_p(x)} \tag{3}$$

Equation (3) has the dimension temperature per time. The parameter $\alpha$ is the temperature conductivity $\alpha = \lambda/\rho c_p$ . The last term reflects the heating rate caused by the absorbed radiant power at point x. In [22] the equation of thermal conductivity is solved numerically. Here we present an analytical solution. Two assumptions must be met for the calculation (Fig. 2):

1) The absorption occurs mainly at the surface of the sample. The absorption of the glass substrate can be neglected.

2) The solution is valid only after a short time after the heat pulse (i.e. the temperature on the backside of the sample ($x=d_s$) is still very close to the initial temperature $T_0$).

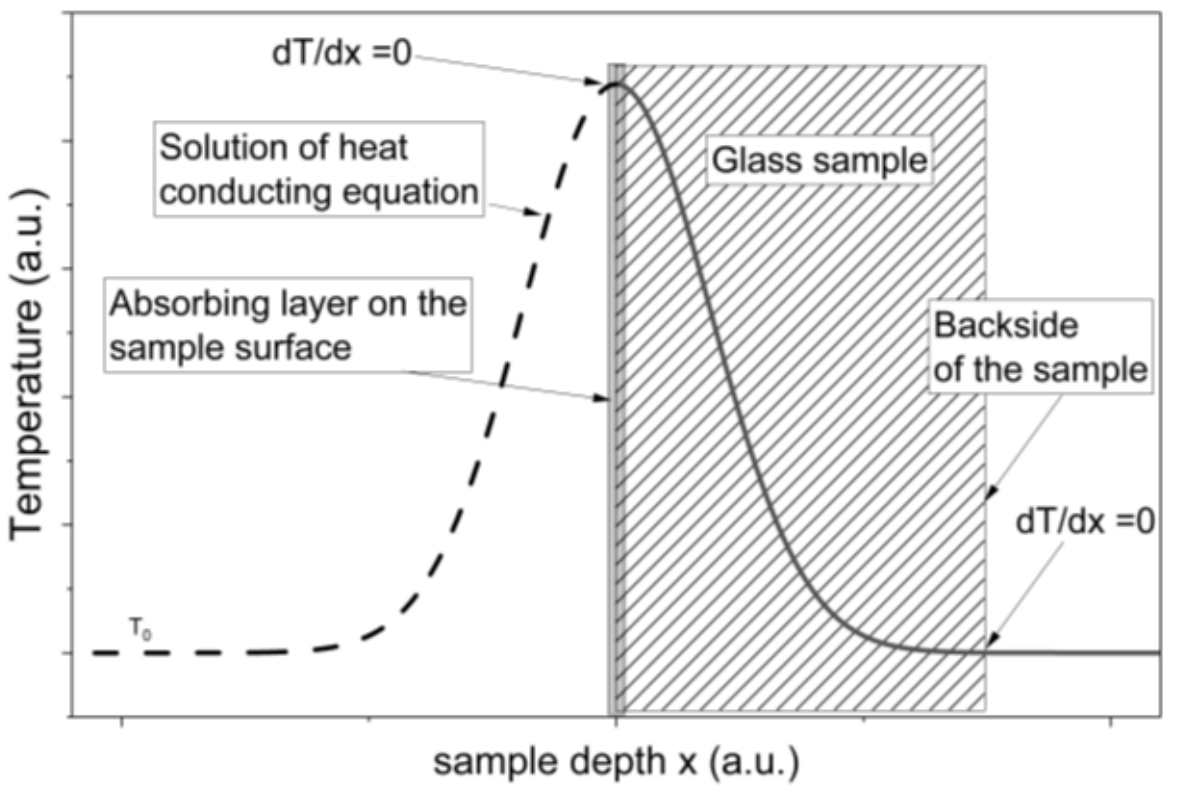

Fig. 2. Temperature gradient depending on the substrate thickness

If the assumptions illustrated in **Fig. 2** are valid the differential equation (3) can be solved for an infinite space. This allows the application of fundamental solution

$$H\ (x,t) = \frac{1}{\sqrt{4\pi\alpha t}} e^{-\frac{x^2}{4\alpha t}} \tag{4}$$

If the temperature is $T_0$ in the entire specimen at the time $t_0$ then one can obtain the solution

$$T(x,t) = T_0 + \int_{t_0}^{t} \frac{1}{\sqrt{4\pi\alpha(t-t')}} \int_{0}^{d_S} e^{-\frac{(x-x')^2}{4\alpha(t-t')}} \frac{q_{vol}((x-x'),(t-t'))}{c_v(x-x')} dx'dt' \tag{5}$$

Since the absorption in the glass is negligible compared to the absorption in the layer, and since the layer has been positioned at x = 0, the source term can be written as $q_{vol} = Q_{vol}\delta(x')$ . This simplifies the formula (5) to

$$T(x,t) = T_0 + \int_{0}^{t} \frac{1}{\sqrt{4\pi\alpha(t-t')}} e^{-\frac{x^2}{4\alpha(t-t')}} \frac{Q_{vol}(t-t')}{c_v} dt' \tag{6}$$

The formula was implemented in a Java script. The calibration was done based on experimental data of the absorbed radiation energy.

3. Results and discussion

3.1. Thermal models of FLA annealing

FLA is a specific technique characterized by delivering a large amount of energy into the film within a short timeframe. This energy is converted into heat, which aids in reducing defects and stress within the film. To gain a better understanding, the impact of FLA on the surface temperature of individual films was modeled. Greater heat transfer occurs in films with a higher nitrogen content, primarily due to their lower optical transmittance. This trend is observed independently of the delivered energy. The maximum estimated temperatures during FLA are listed in **Table 1**, where this trend is clearly demonstrated. While the nitrogen-free film reaches a maximum temperature of approximately 767 °C, the film with the maximum nitrogen content (90 % nitrogen fraction) reaches a maximum temperature of around 905 °C.

Table 1. Maximum surface temperature for different $E_{density}$ and different $f_{N2}$

| $E_{density}$ | Maximum surface temperature (° C) | | |
|---|---|---|---|
| $Jcm^{-2}$ | $f_{N2}$= 0 % | $f_{N2}$= 10 % | $f_{N2}$= 90 % |
| 4.9 | 283 | 299 | 331 |
| 9.8 | 627 | 666 | 739 |
| 11.7 | 767 | 816 | 905 |

**Fig. 3** illustrates the effect of pulse energy for two extreme cases: the lowest delivered energy density of 4.9 $Jcm^{-2}$ and the highest of 11.7 $Jcm^{-2}$. An important parameter for investigating the effect of FLA is the duration for which the film was heated above the deposition temperature (200 °C). As can be seen from **Fig. 3**, for the films exposed to the lowest energy density of 4.9 $Jcm^{-2}$, the time during which the film remained heated above 200 °C is less than 5 ms for all nitrogen contents. Conversely, for the films exposed to the highest energy density of 11.7 $Jcm^{-2}$, the dwell time above 200 °C exceeds 25 ms. These films have sufficient time and energy to achieve optimal relaxation of the crystal lattice and to eliminate the stress induced during deposition. The effect of the supplied heat on these properties will be further investigated in the following sections, using primarily Raman, XRD, and SEM methods.

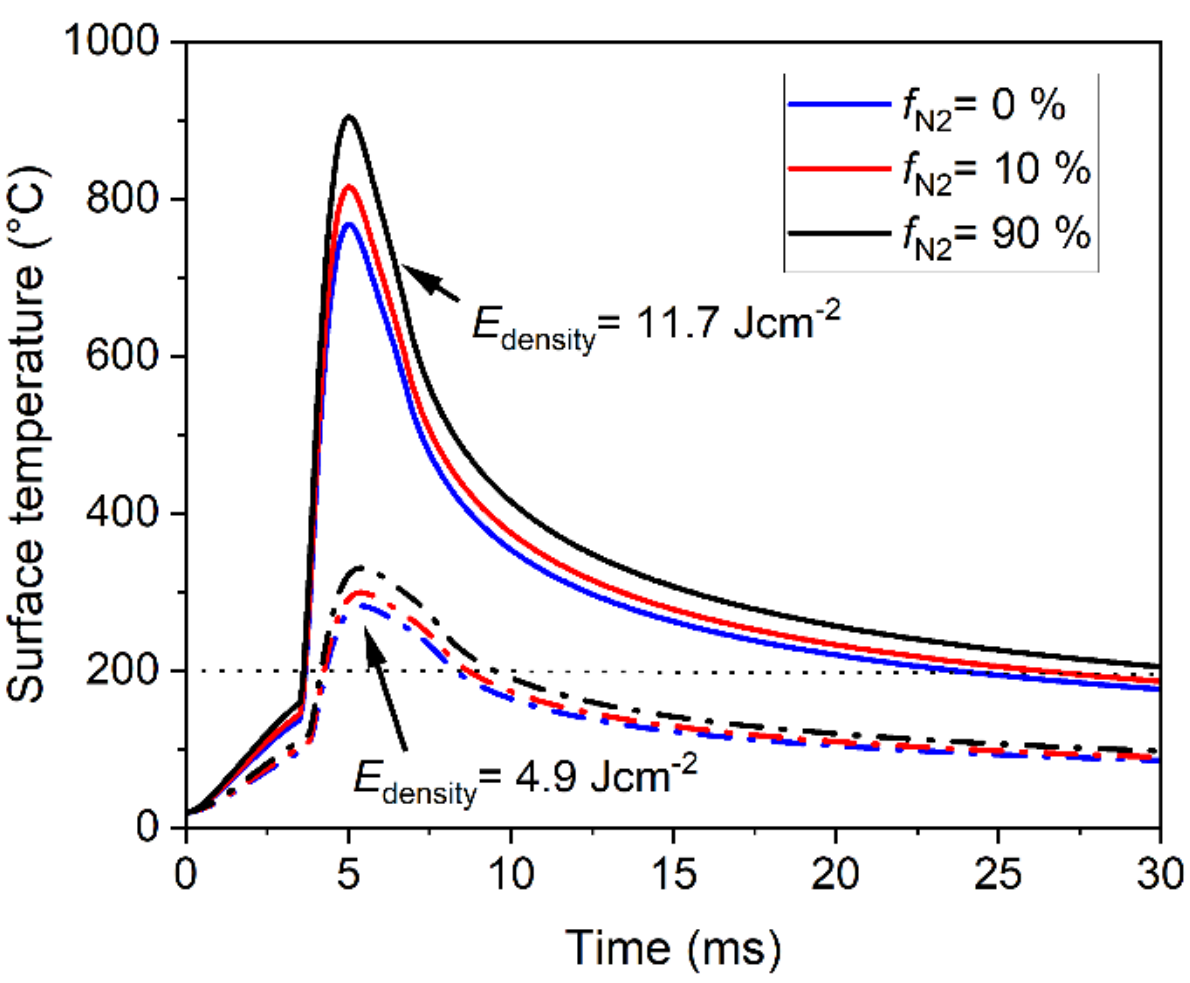

Fig. 3. Calculated surface temperature as a function of time during FLA for different values of $f_{N2}$.

### 3.2. Elemental composition and structural properties

The elemental composition of the $Cu_2O$:N thin films, as determined by WDS, indicates that FLA has a minimal effect on nitrogen desorption. The copper content remains constant at approximately 66 at. % (± 1 at. %) for all as-deposited and FLA-treated films. For the films deposited at $f_{N2}$ = 0 %, the oxygen content is around 33 at. % (± 2 at. %), and no atomic nitrogen is detected in these films after FLA treatment. For the films prepared with $f_{N2}$ = 10%, a nitrogen content of approximately 0.5 at. % (± 20 % rel.) and an oxygen content of 32.5 at. % (± 2 at. %) were observed. The films deposited at $f_{N2}$ = 90 % contain around 2.8 at. % of nitrogen (± 20 % rel.). A slight decrease in the atomic concentration of nitrogen can be observed with increasing FLA pulse energy. However, the differences remain within the experimental error margins. Overall, the FLA treatment has no significant impact on the elemental composition of the films.

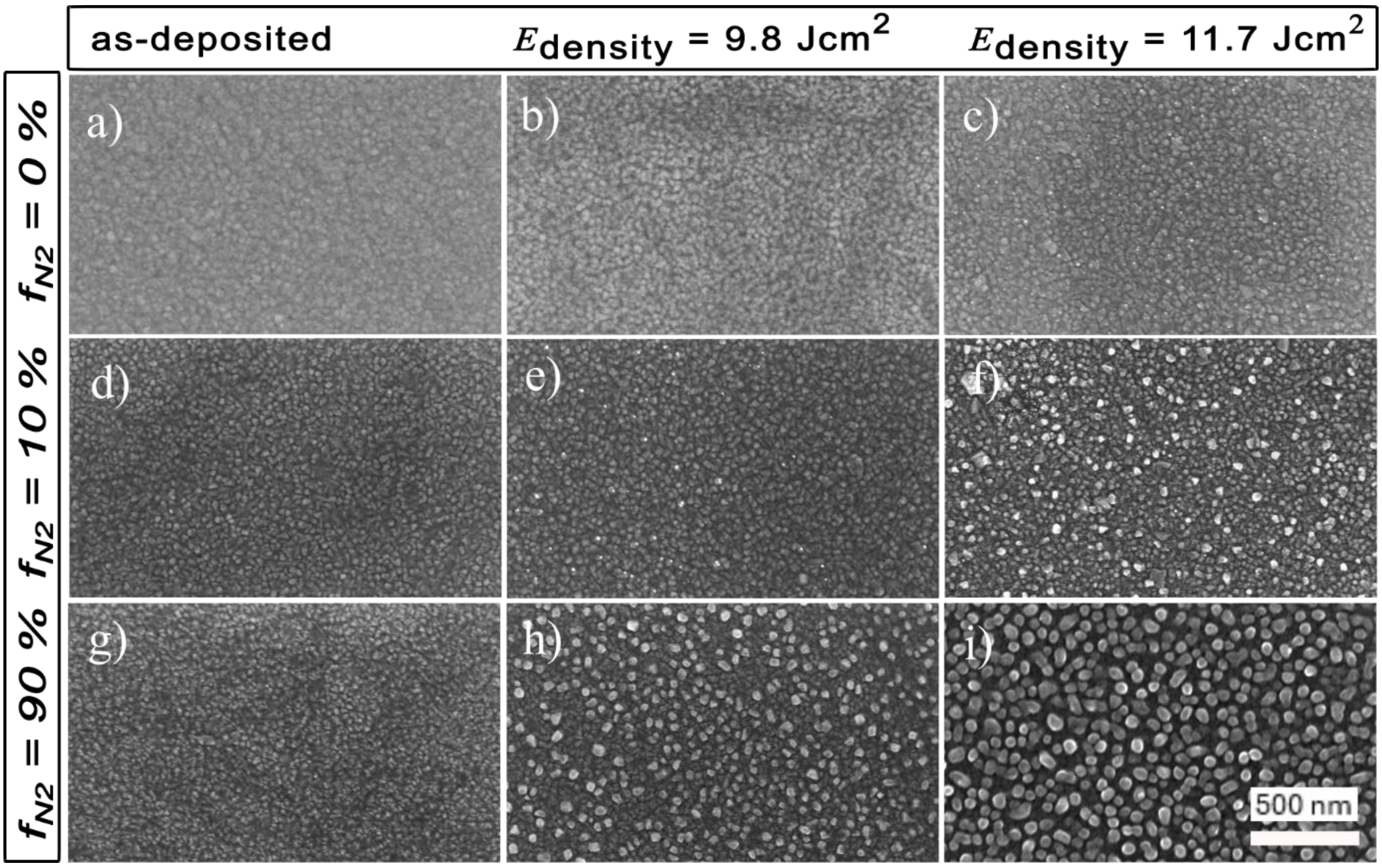

Fig. 4. Top-view SEM micrographs of $Cu_2O$:N films deposited at $f_{N_2}$ = 0, 10, and 90% (top, middle, and bottom rows, respectively). The columns show the as-deposited films (a, d, g) and the films after FLA at energy densities of 9.8 $Jcm^{-2}$ (b, e, h) and 11.7 J $cm^{-2}$ (c, f, i).

Top-view SEM images (**Fig. 4**) illustrate the effect of nitrogen and FLA pulse energy on the surface structure of the films. It is evident that for the films deposited at $f_{N2}$ = 0 %, no significant change in the surface morphology occurs. Furthermore, there is no noticeable difference between the as-deposited films with and without nitrogen. A distinct change in the surface structure occurs only for the nitrogen-containing films. With increasing $E_{density}$, grain growth and coalescence are observed. This trend is more pronounced for the films with higher nitrogen contents.

The XRD patterns in **Fig. 5** confirm that all films retain the cubic $Cu_2O$ phase after FLA, with the principal $Cu_2O$ (111) and $Cu_2O$ (200) reflections located close to their reference positions at 36.418° and 42.297°, respectively. No additional reflections corresponding to CuO or $Cu_4O_3$ were detected. The weak feature at approximately 38.19° originates from the Au contacts deposited on the samples. Changes in the relative intensities and widths of the $Cu_2O$ reflections indicate that both nitrogen incorporation and FLA affect the films' preferred orientation and crystalline order. With increasing FLA energy density, the $Cu_2O$ reflections generally shift toward higher diffraction angles, indicating changes in the interplanar spacing and residual lattice strain rather than a phase transformation. The most pronounced peak sharpening is observed for the film deposited at $f_{N_2}$ = 90% and treated at $E_{density}$ =11.7 J cm$^{-2}$, which is consistent with improved crystallinity and possibly larger coherent scattering domains,

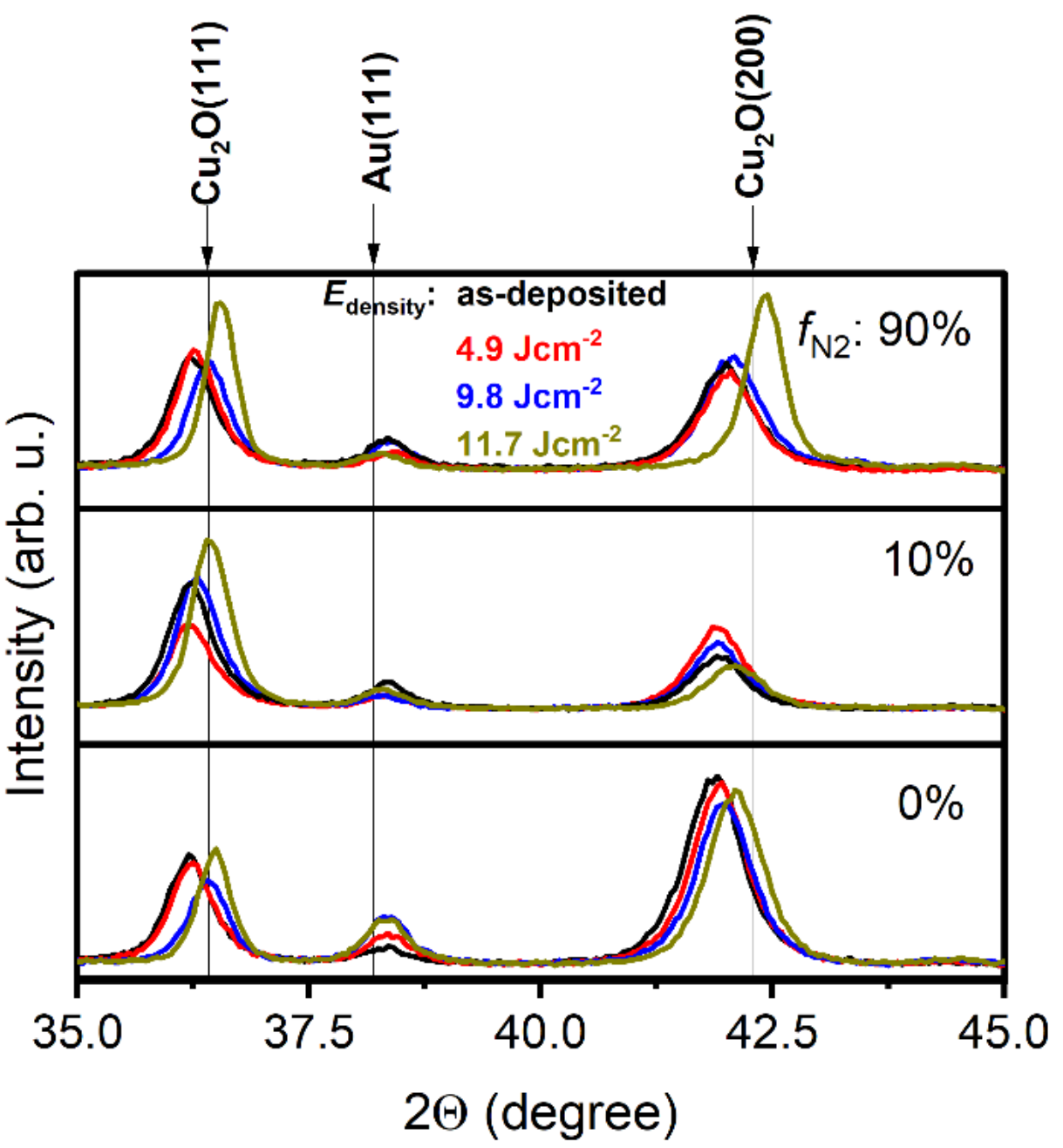

Fig. 5. XRD patterns of the $Cu_2O$:N films at different $E_{density}$ during FLA and different $f_{N2}$

as indicated also in SEM images (**Fig. 4**).

The broad Raman band centred at approximately 2250 cm$^{-1}$ (**Fig. 6**) is attributed to the N–N stretching vibration of molecular nitrogen incorporated in the $Cu_2O$ lattice. Its position agrees with the nitrogen-induced Raman band previously reported at 2250–2270 cm$^{-1}$ and with the calculated vibrational mode near 2280 cm$^{-1}$ for an $N_2$ molecule occupying a Cu site, denoted as $(N_2)_{Cu}$ [24–26]. This assignment is further supported by electron energy-loss spectroscopy, which identified nitrogen predominantly in a molecular rather than an anionic form in sputtered $Cu_2O$:N films [25]. The $(N_2)_{Cu}$ configuration has also been predicted to form a relatively shallow acceptor state and may therefore contribute to the enhanced p-type conductivity of nitrogen-doped $Cu_2O$ [24]. Previous results for reactively sputtered $Cu_2O$:N showed that the intensity of this Raman band can vary substantially with the deposition conditions even when the total nitrogen concentration remains comparable [17]. Consequently, its intensity should be regarded as an indicator of the population and local environment of Raman-active molecular $N_2$ configurations rather than as a direct measure of either the total nitrogen content or the concentration of electrically active acceptors.

For the films deposited at $f_{N2}$ = 10%, the molecular-$N_2$ band generally becomes more pronounced with increasing FLA energy density (**Fig. 6(a)**). This evolution may indicate that the short thermal pulse promotes a redistribution of incorporated nitrogen into Raman-active molecular configurations or modifies the local bonding and polarizability of the existing $N_2$ species. A different, non-monotonic response is observed at $f_{N2}$ = 90% (**Fig. 6(b)**). The initially strong $N_2$ band remains prominent up to intermediate FLA energy densities but weakens considerably at 11.7 J $cm^{-2}$. Because WDS shows no statistically significant decrease in total nitrogen content, the loss of Raman intensity alone does not demonstrate $N_2$ desorption. Instead, it may reflect a redistribution of nitrogen among different lattice environments, conversion to Raman-inactive configurations, or a change in the Raman scattering efficiency caused by the accompanying structural changes of the $Cu_2O$ matrix. Moreover, for $f_{N2}$ = 10%, the increase in the Raman signal at high FLA energies is accompanied by a decrease in the hole concentration (see next section), showing that the Raman-band intensity does not directly track the electrically active acceptor density. For $f_{N2}$ = 90%, the weakening of the band coincides with a marked increase in resistivity. However, in the absence of reliable Hall-effect data for these films, this correlation cannot be assigned to a specific microscopic mechanism

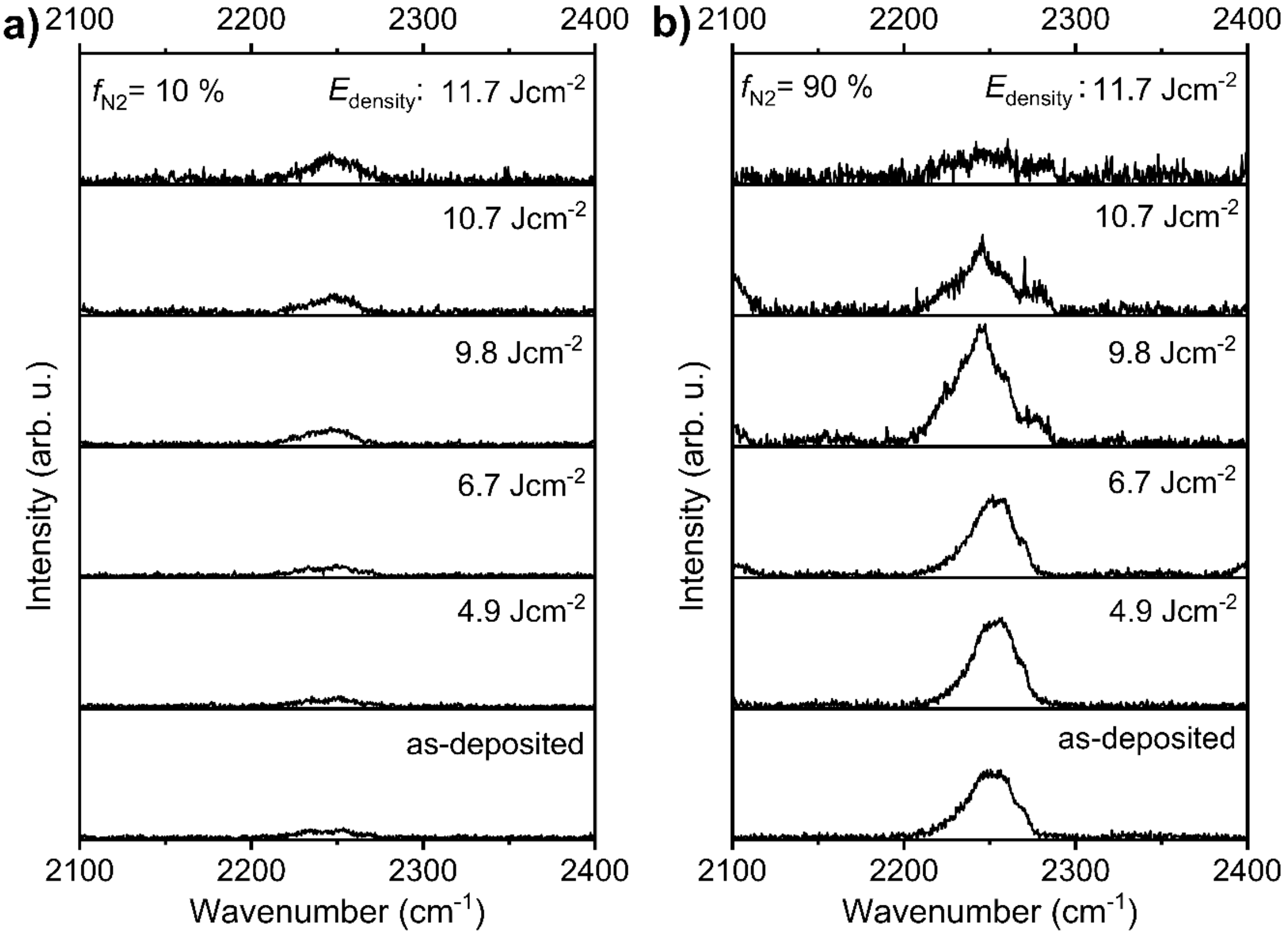

Fig. 6. Raman spectrum of $Cu_2O$:N films prepared at $f_{N2}$ = 10 % (a) and 90 % (b) as a function of energy density during FLA.

### 3.3. Electrical and optical properties

**Fig. 7** shows that nitrogen incorporation significantly decreases the as-deposited resistivity, from approximately $10^2$ Ω cm for the nitrogen-free films to values below $10^0$ Ω cm at the highest nitrogen contents. The response to FLA, however, strongly depends on both $f_{N2}$ and the applied energy density. For $f_{N2}$ = 0%, the resistivity remains on the order of $10^2$ Ω cm and exhibits no systematic improvement after FLA (**Fig. 7 (a)**). The films deposited at $f_{N2}$ = 10% remain relatively stable at 4.9 and 6.7 J cm$^{-2}$, whereas their resistivity begins to increase at 9.8 J cm$^{-2}$ and continues to deteriorate at higher energy densities (**Fig. 7 (b)**). At $f_{N_2}$ = 40%, FLA at 4.9 and 6.7 J cm$^{-2}$ produces a modest reduction in resistivity, followed by a progressive increase at higher energy densities (**Fig. 7 (c)**). The strongest and most clearly non-monotonic response is observed for $f_{N2}$ = 90% (**Fig. 7 (d)**). The minimum resistivity of $4.5 \times 10^{-2}$ Ω cm is obtained after FLA at 4.9 J cm$^{-2}$, which is lower than the value reported in our recent work [17]. The resistivity remains below the corresponding as-deposited values up to 8.2 J $cm^{-2}$ but increases sharply from 9.8 J $cm^{-2}$ and reaches several hundred Ωcm at 11.7 J $cm^{-2}$. At the highest energy density, the conductivity advantage provided by nitrogen incorporation is therefore almost completely lost, and the resistivity approaches that of the nitrogen-free films. The onset of this degradation occurs in the same energy range as the non-monotonic evolution of the molecular-$N_2$ Raman band, which weakens substantially at 11.7 J $cm^{-2}$ (**Fig. 6b**). Although this correlation suggests that changes in the local nitrogen configuration may contribute to the electrical degradation, the Raman data alone do not establish a causal mechanism. Overall, low-energy FLA provides a limited processing window for improving the conductivity of nitrogen-rich $Cu_2O$:N films, whereas excessive energy density leads to a pronounced deterioration of their electrical properties.

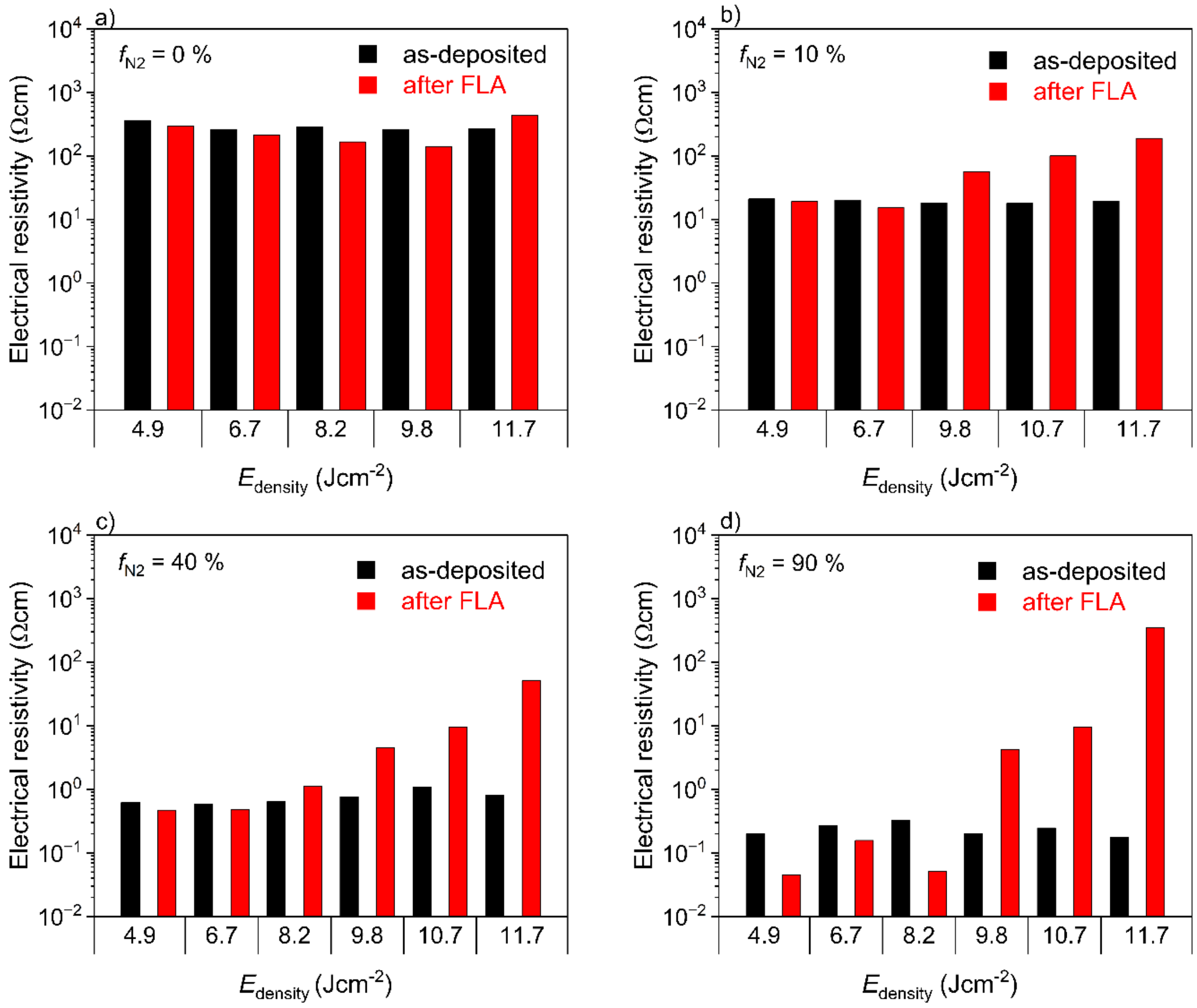

Fig. 7. Electrical resistivity as a function of energy density during FLA of the $Cu_2O$:N films prepared without nitrogen (a), at $f_{N2}$ = 10 % (b), at $f_{N2}$ = 40 % (c), and $f_{N2}$ = 90 % (d).

To clarify the origin of the resistivity changes, Hall-effect measurements were performed. Because the electrical resistivity is given by $\rho = (q\ p\ \mu_h)^{-1}$, where $p$ is the hole concentration and $\mu_h$ is the hole mobility, the measured resistivity reflects the combined evolution of both parameters. Reliable Hall data could be obtained only for the films deposited at $f_{N2}$ = 0% and 10%. For the films prepared at higher $f_{N2}$, the Hall signal was below the reliable measurement limit, most likely because of the combination of very low mobility and high carrier concentration.

In the as-deposited state, the nitrogen-free films exhibit a mobility on the order of $10^0$ cm$^2$ V$^{-1}$ s$^{-1}$ and a hole concentration on the order of $10^{16}$ cm$^{-3}$. In contrast, the films deposited at $f_{N2}$ = 10% exhibit a substantially lower mobility, on the order of $10^{-2}$ cm$^2$ V$^{-1}$ s$^{-1}$, but a much higher hole concentration, on the order of $10^{18}$ cm$^{-3}$. The reduction in resistivity induced by nitrogen incorporation is therefore primarily associated with the increase in hole concentration, which more than compensates for the accompanying decrease in mobility.

As shown in **Fig. 8**, increasing the FLA energy density generally increases the hole mobility while decreasing the hole concentration. For the nitrogen-free films, mobility increases at intermediate energy densities, reaching a maximum of 7.6 cm$^2$ V$^{-1}$ s$^{-1}$ at 9.8 J cm$^{-2}$. This mobility improvement is accompanied by a moderate decrease in hole concentration, so the resulting change in resistivity remains comparatively small. The effect is considerably more pronounced for the films deposited at $f_{N2}$ = 10%. Above 9.8 J cm$^{-2}$, the mobility increases progressively, whereas the hole concentration decreases by more than one order of magnitude. The decrease in $p$ compensates the improvement in $\mu_h$, reducing the product $p\mu_h$ and thereby explaining the marked increase in resistivity observed in **Fig. 7(b)**.

The onset of these transport changes coincides with the growth of the Raman band assigned to molecular $N_2$. Together with the shifts observed in the XRD patterns, these results suggest that high-energy FLA alters the local lattice and defect structure, thereby affecting both carrier scattering and the population of electrically active defects.

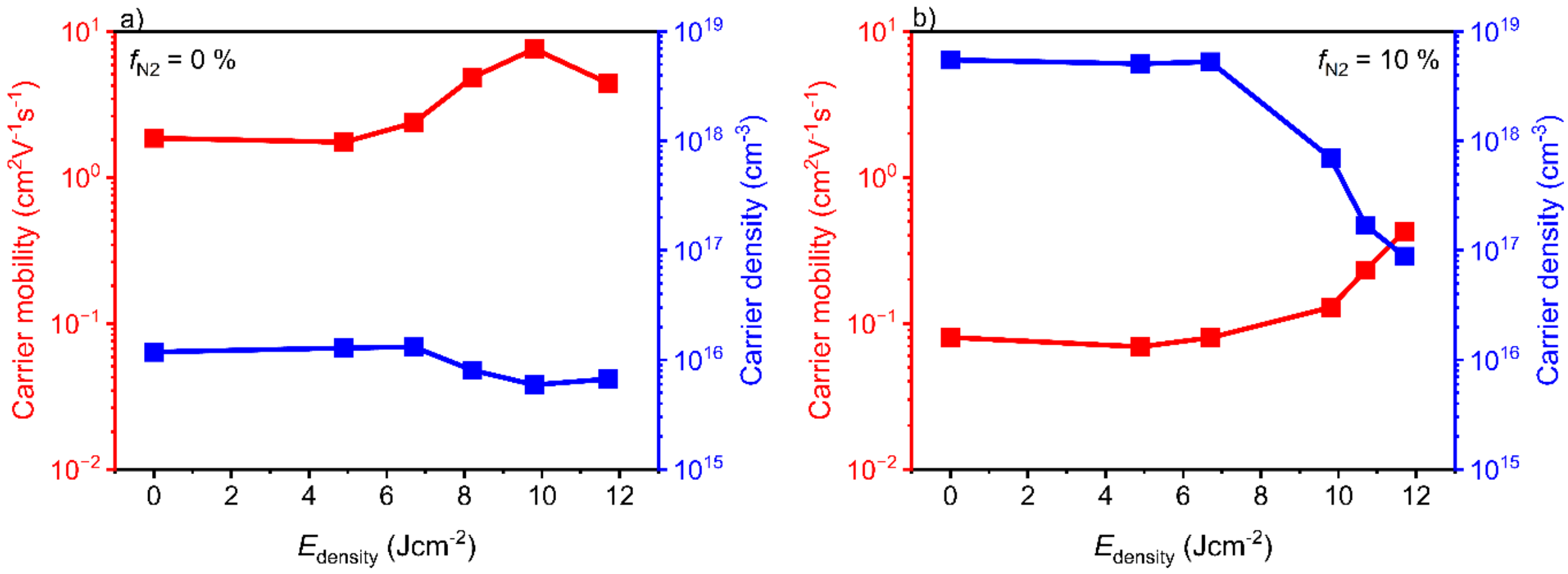

Fig. 8. Carrier mobility and carrier density as a function of energy density during FLA of the Cu-O-N films prepared at $f_{N2}$ = 0 % (a) and 10 % (b).

The optical band-gap values in **Fig. 9** show a clear dependence on nitrogen incorporation. In the as-deposited state, the band gap decreases progressively from approximately 2.54 eV for $f_{N2}$ = 0% to approximately 2.43 eV for $f_{N2}$ = 90%. This red shift suggests that nitrogen incorporation modifies the electronic states near the band edges. Possible contributions include nitrogen-related defect states, increased structural disorder, and changes in lattice strain. Because the band gap was extracted from the optical absorption edge, the observed shift may also be influenced by defect-induced band tailing and should not automatically be interpreted as a change in the intrinsic band structure alone.

FLA has a negligible effect on the optical band gap of the films deposited at $f_{N2}$ = 0% and 10%, with the measured values remaining nearly constant over the investigated energy range. In contrast, the films deposited at $f_{N2}$ = 40% and 90% exhibit a non-monotonic response. At lower energy densities, only small variations are observed, whereas high-energy FLA produces pronounced band-gap widening. The strongest change occurs for $f_{N2}$ = 90%, for which the band gap increases from approximately 2.43 eV in the as-deposited state to approximately 2.54 eV after FLA at 11.7 J $cm^{-2}$.

This composition-dependent response indicates that the electronic structure of the nitrogen-rich films is more sensitive to the rapid thermal treatment. The band-gap widening at high energy densities may be associated with changes in crystalline ordering, residual lattice strain, or the density and energetic distribution of defect-related states near the band edges, as indicated GIXRD and SEM analysis.

Fig. 9. Optical band gap as a function of $E_{density}$ during FLA.

## 4. Conclusions

This study demonstrates that millisecond FLA is an effective tool for tailoring the properties of reactively sputtered $Cu_2O$:N films, with the resulting response governed by nitrogen incorporation and pulse energy density. WDS revealed no statistically significant changes in the total elemental composition, while GIXRD confirmed that all films retained the cubic $Cu_2O$ phase without detectable secondary copper oxides. Nitrogen-containing films exhibited surface coarsening and changes in lattice spacing, residual strain, and crystalline ordering. The evolution of the Raman band near 2250 $cm^{-1}$ further indicates that FLA modifies the local environment of molecular $N_2$ without producing a measurable loss of total nitrogen. Nitrogen incorporation substantially reduced the resistivity of the as-deposited films, and the lowest value of $4.5 \times 10^{-2}$ Ω cm was obtained for $f_{N2}$ = 90% after FLA at 4.9 J $cm^{-2}$. Hall measurements for $f_{N2}$ = 0% and 10% showed that FLA increased hole mobility while decreasing hole concentration, demonstrating that the electrical response results from a balance between these two parameters. Nitrogen incorporation also reduced the as-deposited optical band gap, whereas higher FLA energy densities widened the extracted band gap of nitrogen-rich films, most likely through changes in crystalline ordering, strain, and defect-related absorption.

Overall, the results identify a well-defined low-energy processing window in which FLA enhances the electrical properties of nitrogen-rich $Cu_2O$:N while preserving its phase and composition. By selecting the appropriate pulse energy density, FLA therefore offers a rapid and promising route for tuning the structural, electrical, and optical properties of p-type $Cu_2O$:N thin films for various optoelectronic applications.

**CRediT authorship contribution statement**

**Jan Koloros:** Investigation, Methodology, Visualization, Writing – original draft Writing – review & editing. **Pavel Baroch:** Supervising, Conceptualization. **Thomas Preußner**: Investigation, Methodology. **Matthias Fahland**: Investigation, Methodology **Michaela Červená:** Visualization. **Jiří Rezek**: Conceptualization, Methodology, Investigation, Visualization, Writing – original draft, Writing – review & editing.

**Acknowledgement**

This work was supported by the project Quantum materials for applications in sustainable technologies (QM4ST), funded as project No. CZ.02.01.01/00/22_008/0004572 by Programme Johannes Amos Comenius, call Excellent Research.

**Declaration of generative AI and AI-assisted technologies**

During the preparation of this work, the author used ChatGPT and Grammarly for state-of-the-art research, to improve the clarity of the text, and for grammar checking. The author reviewed and edited the output as needed and takes full responsibility for the content of the published article.